\documentclass[12pt]{article}

\usepackage{geometry,graphicx}

\usepackage{pslatex}
\usepackage{hyperref} 

\usepackage{graphicx}
\begin{document}
\begin{center}                  
{\large\bf Observation of the $\Xi_b^0$ baryon } \vspace*{3.0mm}\\ 
Peter J Bussey.\\[3mm]
University of Glasgow, Glasgow G12 8QQ, U.K.\\
E-mail: peter.bussey@glasgow.ac.uk \vspace{2mm}\\
{\em for the CDF Collaboration}

\end{center}

\abstract{The first observation of the heavy baryonic state $\Xi_b^0$
is reported by the CDF Collaboration. A new decay mode of the
established state $\Xi_b^-$ is also observed.  In both cases the decay
into a $\Xi_c$ plus a charged pion is seen, with an equivalent
statistical significance of above $6.8\sigma$.}

~\\[5mm]
Presented at the 2011 Europhysics Conference on High Energy Physics, EPS-HEP 2011,\\
		July 21-27, 2011\\
		Grenoble, Rh\^one-Alpes, France.

\section{Introduction}
The quark model of elementary particles is well established and has a
impressive history of success in its account of hadronic
states. Nevertheless, it is important to continue to test it by
searching for hitherto unobserved particles that are predicted to
exist, both to provide continued confirmation of the quark model, and
to provide a background for the possible discovery of unusual types of
particle.  In this presentation we report the first observation, by
the CDF Collaboration, of a new baryonic state, the
$\Xi_b^0$~\cite{cdfxib0}. This consists of a $\,b\,s\, u$ quark
combination and fills an important gap in the set of baryons containing
a $b$ quark.

One possible search strategy would be to look for decays of a
hypothesized $b$ state into the easily-identifiable $J/\psi$ meson.
This will not succeed in the present case because of the unavoidable
presence of a $\pi^0$ in the decay chain, which is hard to vertex
precisely. Instead, we have made use of the decay
$\Xi_b^0\to\Xi_c^+\pi^-$, which has the merit that the previously
unobserved decay $\Xi_b^-\to\Xi_c^0\pi^-$ can also be sought using a
parallel search scheme.  The $\Xi_b^-$ is a known state, and its
observation provides a check on the methodology employed here.

The decay chains that are used in the present study are as follows.
(Throughout this account, the corresponding charge conjugate states are
also implied.)
\begin{center}
\begin{tabular}{rclrcl}
$\Xi_b^0$ & $\to$ & $\Xi_c^+\,\pi^-$ & \hspace*{0.1\textwidth} 
$\Xi_b^-$ & $\to$ & $\Xi_c^0\,\pi^-$ \\
$\Xi_c^+$ & $\to$ & $\Xi^-\,\pi^+\pi^+$ & 
$\Xi_c^0$ & $\to$ & $\Xi^-\,\pi^+$  \\
$\Xi^-$ & $\to$ & $\Lambda\,\pi^-$ & $\Xi^-$ & $\to$ & $\Lambda\,\pi^-$ \\
$\Lambda$ & $\to$ & $p\,\pi^-$ & $\Lambda$ & $\to$ & $p\,\pi^-$ 
\end{tabular}
\end{center} 
The $\Xi_b^0$ decay chain is depicted in figure 1.

\begin{figure}[!t]
~\\[-15mm]
\begin{center}
\includegraphics[width=.5\textwidth]{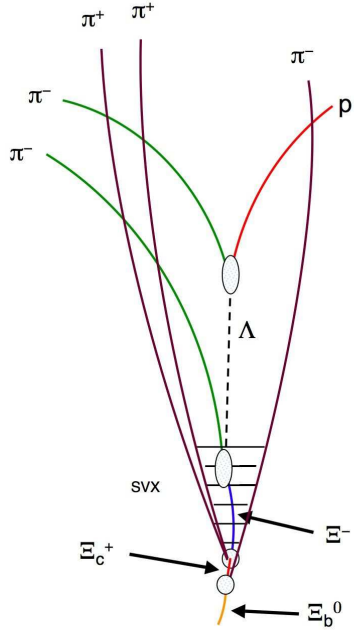}
~\\[-10mm]
\caption{ Schematic depiction of the $\Xi_b^0$ decay chain in the mode
studied here.}
\end{center}
\end{figure}

\section{Search method}

The CDF II detector is a multi-purpose detector situated at the
Fermilab Tevatron.  Inside a solenoid that provides a 1.4 T axial
magnetic field, an array of coaxial silicon detector layers (SVX) is
surrounded by the Central Outer Tracker, which is a drift detector containing 
axial and angled layers of wires.  It is these detector systems that
are used in the present analysis. A displaced track trigger, using
silicon and COT information, is used to record suitable events for
analysis by requiring two charged tracks that do not point to the
beam line. A data sample of integrated luminosity 4.2 fb$^{-1}$ formed
the basis of the present search.

 The reconstruction method starts with an identification of $\Lambda$
candidates by plotting the invariant mass of track pairs of opposite charge
and transverse momentum ($p_T$) greater than 0.4 GeV/c.  The
higher-momentum track is taken as a proton and the lower as a
$\pi^-$.  The tracks must form a good vertex more than 1 cm from the
beamline. Mass cuts around the $\Lambda$ peak select suitable
candidates with good efficiency.

  The $\Lambda$ candidates are then paired with charged tracks to form
$\Xi^-$ candidates, taking the charged track as a $\pi^-$ and requiring a
good quality $\Lambda\pi^-$ intersection at a vertex point. The flight
distance of the $\Lambda$ from the $\Xi^-$ vertex and the flight
distance of the latter from the beamline must both exceed 1 cm.
Silicon detector hits are now demanded on the postulated $\Xi^-$ track, and a
new vertex fit is performed. This procedure improves the quality of
the $\Xi^-$ peak-to-background ratio dramatically, as seen in figure
2.

\begin{figure}[!b]
\begin{center}
\includegraphics[width=.4\textwidth]{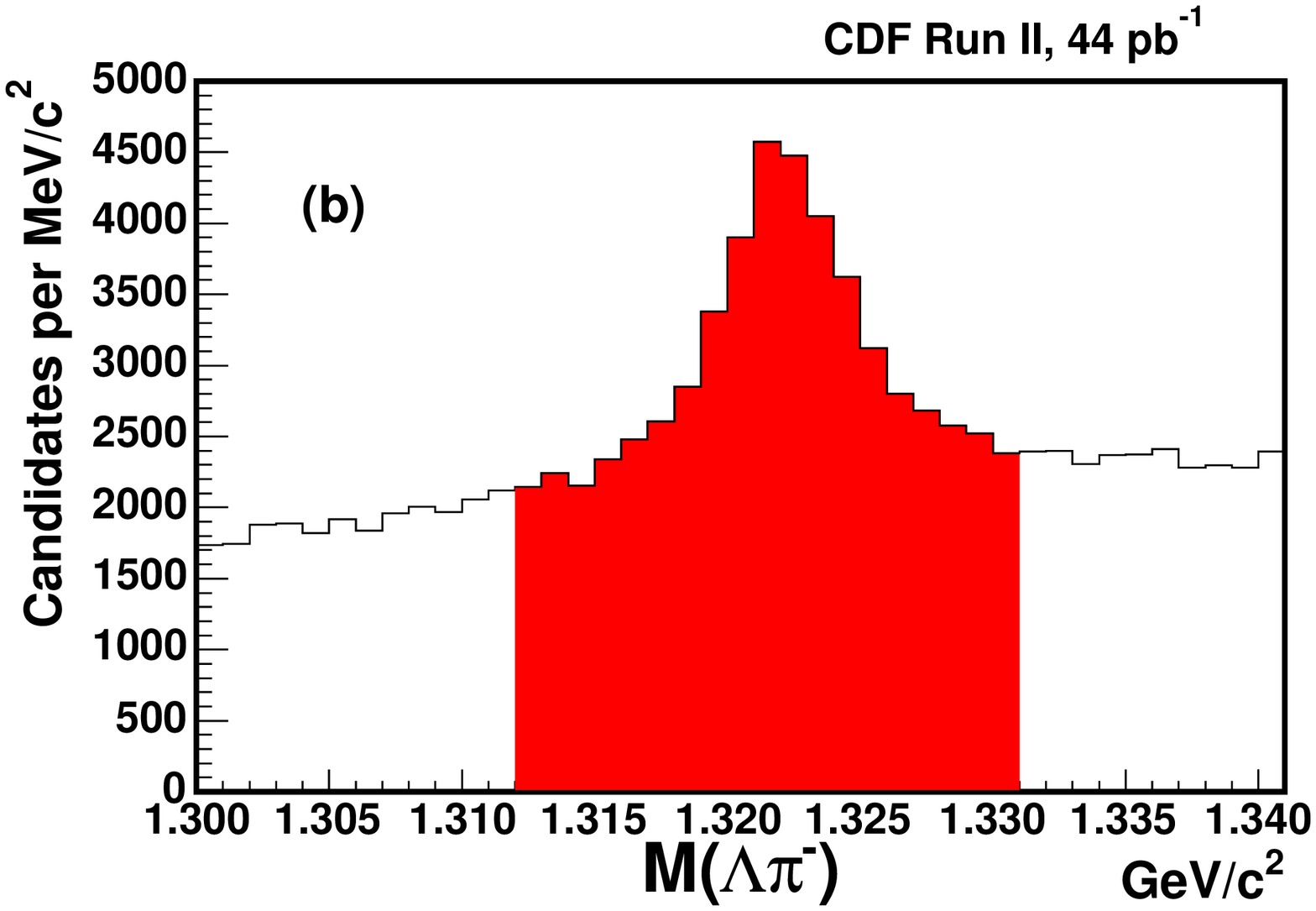}
\includegraphics[width=.4\textwidth]{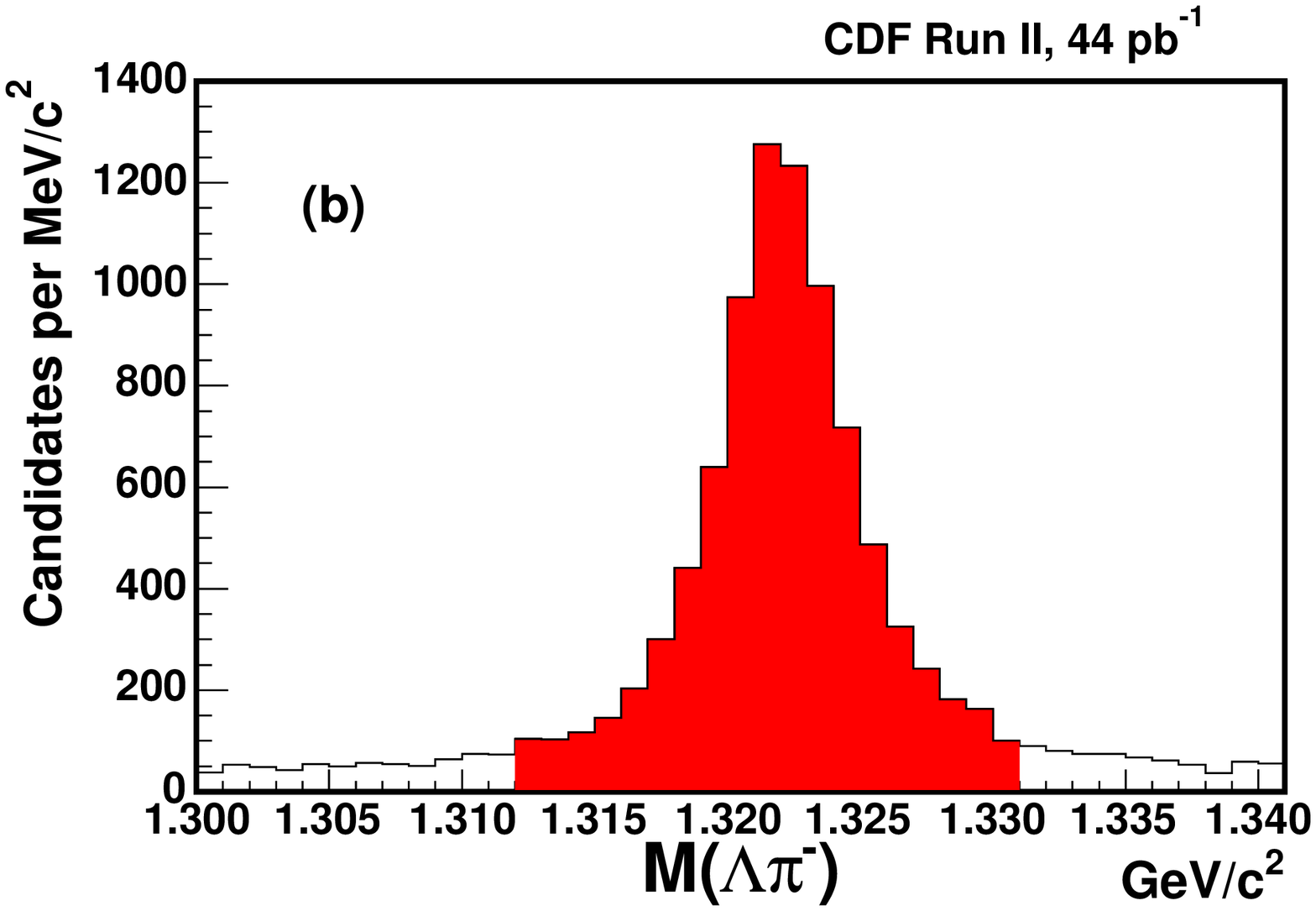}
\caption{Illustrations of $\Lambda\pi^-$ invariant mass distribution (a) without and (b) with the requirement of silicon detector hits on the expected $\Xi^-$ track.}
\end{center}
\end{figure}

The third step in reconstruction is to combine the $\Xi^-$ candidates
with one or two positive tracks, taken as $\pi^+$, to form $\Xi_c$
candidates (figure 3a, b).  The pion tracks must have more than two
silicon detector hits for precise vertexing, a $p_T$ of at least 2
GeV/c, and an impact parameter of at least 100 $\mu$m relative to the
beam line. The combination must have a well-fitted vertex, a total
$p_T$ of at least 4 GeV/c, and a calculated decay $ct$ value of at
least 100 $\mu$m. This procedure yields well-characterised $\Xi_c$
mass peaks at the expected values (figure 3a,b). A selection broadly
covering the peaks is made to maximise efficiency, retaining
2110 $\pm$70 $\Xi_c^0$ candidates and 3048 $\pm$67 $\Xi_c^+$.

\begin{figure}[t]
\begin{center}
\includegraphics[width=.49\textwidth]{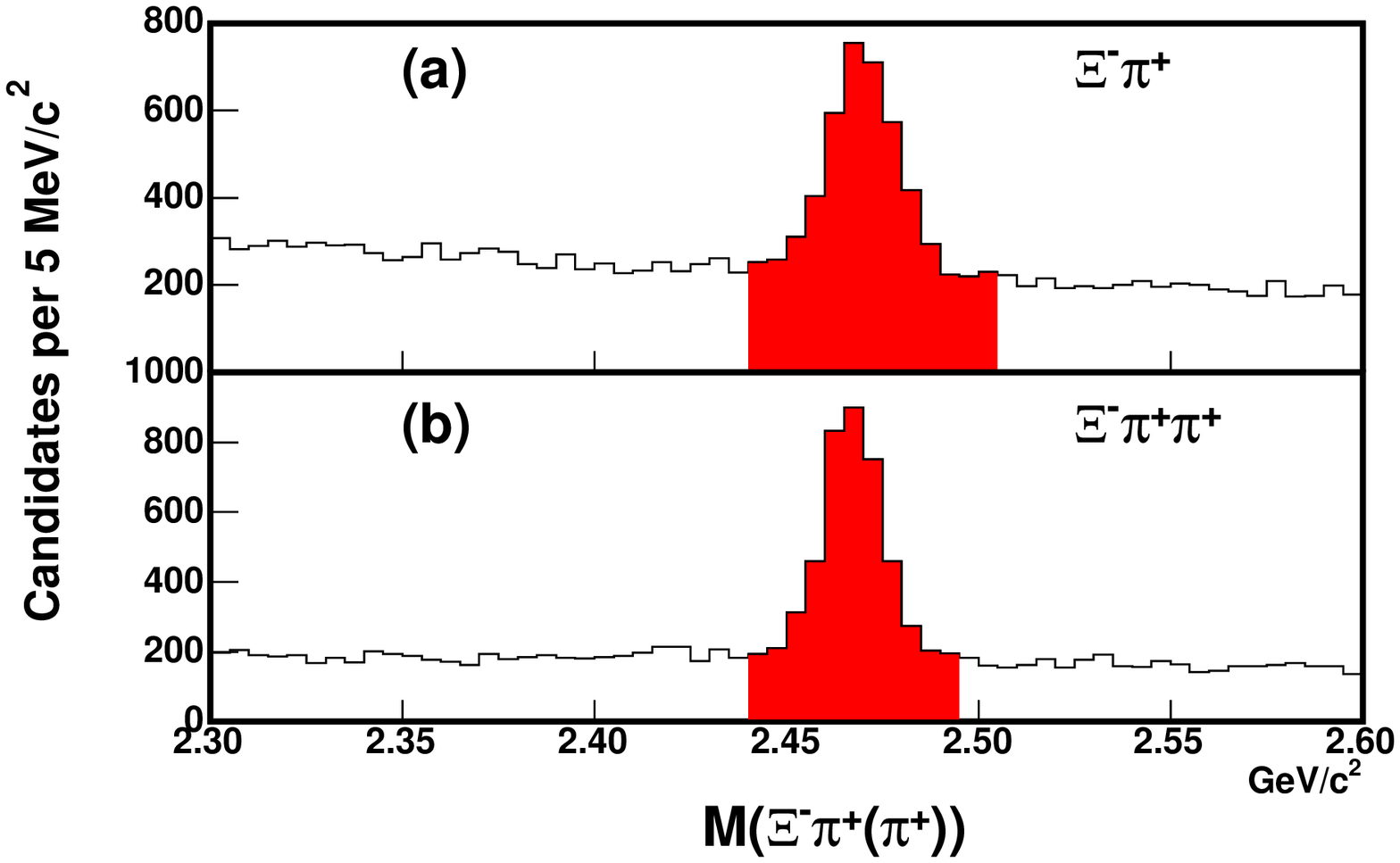}  \\
\includegraphics[width=.49\textwidth]{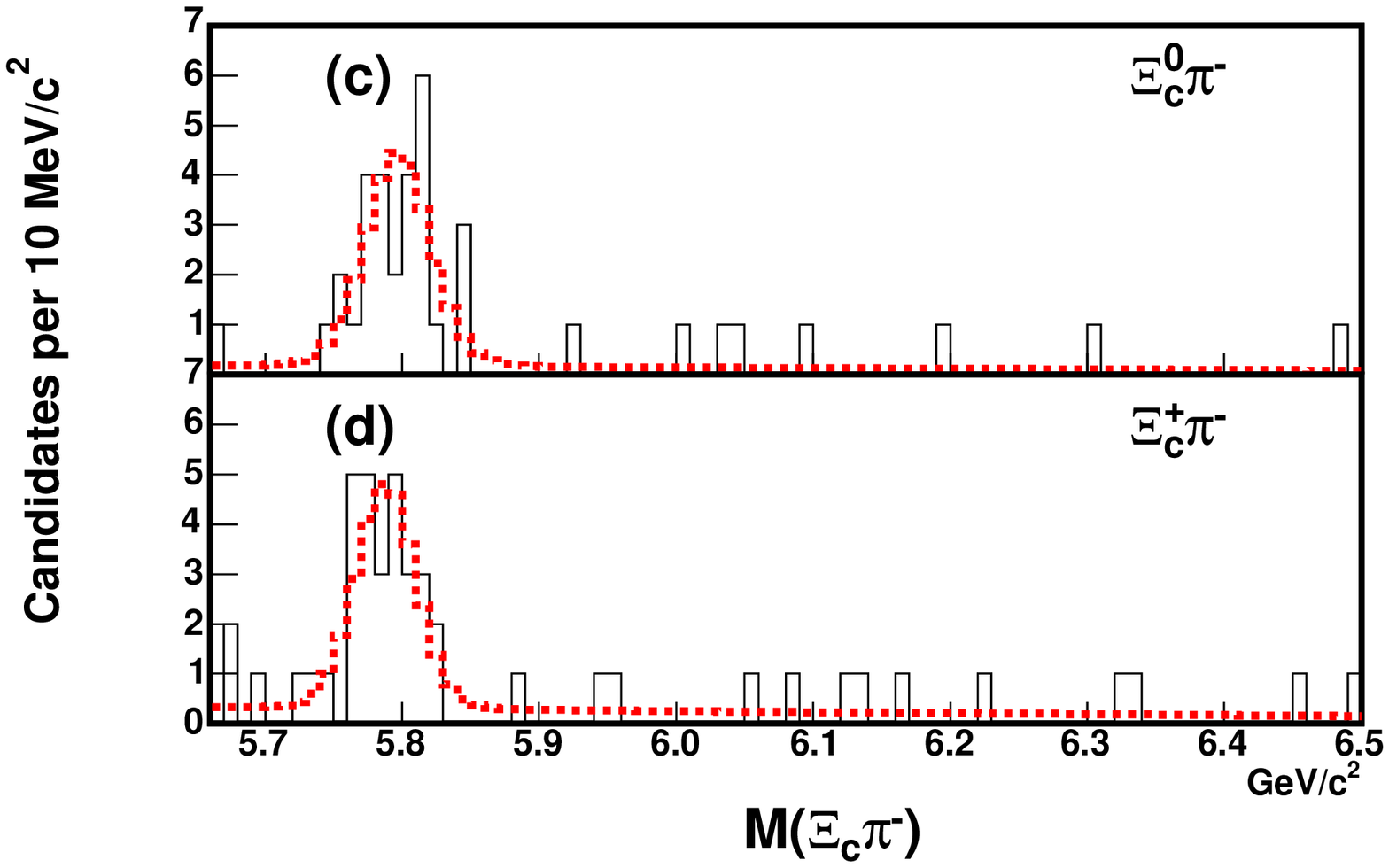}
\caption{(a) $\Xi^-\pi^+$ and (b) $\Xi^-\pi^+\pi^+$ invariant mass
distributions showing $\Xi_c^0$ and $\Xi_c^+$ peaks; \protect \\ (c)
$\Xi_c^0\pi^-$ and (d) $\Xi_c^+\pi^-$ invariant mass distributions,
showing $\Xi_b^-$ and $\Xi_b^0$ peaks.}
\end{center}
\end{figure}

Finally, the $\Xi_c$ candidates are each paired with a further negative
track, taken as a $\pi^-$. This must satisfy $p_T >6$ GeV/c, and a
fully constrained vertex fit is performed using the extrapolated
$\Xi_c$ direction, enforcing all the previously assumed baryon
masses. The position of the fitted vertex must correspond to a
subsequent $\Xi_c$ decay time that satisfies $-2\sigma<t<3t+2\sigma$,
using the measurement uncertainty $\sigma$ and the known mean decay
time $t$ of the $\Xi_c$. This formalism takes account of the very different 
decay distances, 440 $\mu$m and 110 $\mu$m, of the $\Xi_c^+$ and
$\Xi_c^0$ respectively.  At least one of the charged
pions used at this stage and also at the previous stage must satisfy
the displaced track trigger requirements.

\section{Results}

The results of the final stage are shown in figure 3c, 3d, where the
invariant masses of the $\Xi_c\pi^-$ combinations are plotted.  In
both combinations, a clear mass peak is seen.  In the $\Xi_c^0\pi^-$
case the peak is interpreted as the established $\Xi_b^-$ state at
this mass, of which this is an expected decay channel, thus providing
a check on the present procedure. We interpret the peak in the
$\Xi_c^+\pi^-$ combination as an observation of the predicted
$\Xi_b^0$ state, seen here for the first time.  The two peaks are of
similar magnitude, consistent with general expectations.  When an
unbinned likelihood fit is performed to a Gaussian peak on a linear
background, an equivalent significance of 6.8$\sigma$ is found in both
cases using a likelihood ratio test.

Systematic uncertainties are estimated by comparing particle masses
measured in the present fit process with their standard values.  The
fit gives peaks contining 25.8$^{+5.5}_{-5.2}$ $\Xi_b^-$ baryons and
25.3$^{+5.6}_{-5.4}$ $\Xi_b^0$. The fitted mass of the $\Xi_b^0$ is
5787.8 $\pm$5.0 (stat.) $\pm$1.3 (sys.) GeV and that of the $\Xi_b^-$ is
5796.7 $\pm$5.1 (stat.) $\pm$1.4 (sys.) GeV, in agreement with the
value of 5791 $\pm$3 GeV measured in the $J/\psi\,\Xi^-$ decay
channel~\cite{cdfxibm}.  Using the latter $\Xi_b^-$ mass value, we
obtain a mass difference $m(\Xi^-_b)-m(\Xi^0_b)$ between the two states of 3.1
$\pm$5.6 (stat.) $\pm$1.3 (sys.) GeV.

In summary, measuring $\Xi_c\pi^-$ decay modes using the CDF II
detector, we have made the first observation of the $\Xi_b^0$ state,
and also the first observation of the $\Xi_b^-$ state in this decay mode.


\begin{thebibliography}{99}
\bibitem{cdfxib0}  CDF Collab., T. Aaltonen et al., Phys.\ Rev.\ Lett.\ 107 (2011) 031104 [{\tt hep-exp/1107.4015}]. 
\bibitem{cdfxibm}  CDF Collab., T. Aaltonen et al.,  Phys.\ Rev.\ D 80 (2009) 072003 [{\tt hep-exp/0905.3123}].


\end{thebibliography}
\end{document}